\documentstyle[aps,prl,preprint,floats,epsfig]{revtex}

\newcommand{\eeqq}	{\mbox{$e^+e^-\to q\bar q$}}
\newcommand{\KS}{\mbox{$K_S^0$}}
\newcommand{\acp}	{\mbox{${{\cal A}_{\rm CP}}$}}
\newcommand{\cossph}	{\mbox{$\cos \theta_{\rm s}$}}
\newcommand{\dedx}	{\mbox{$dE/dx$}}
\newcommand{\gev}	{\mbox{${\rm ~GeV}$}}
\newcommand{\etapr}	{\mbox{${\eta^\prime}$}}
\newcommand{\fbar}	{\mbox{$\bar f$}}
\newcommand{\branch}    {\mbox{${\cal B}$}}
\newcommand{\fisher}    {\mbox{${\cal F}$}}

\begin{document}

\draft  
\preprint{\tighten\vbox{\hbox{\hfil CLNS 99/1651} 
                        \hbox{\hfil CLEO 99-17}   
}}
\title{Measurement of Charge Asymmetries in Charmless Hadronic
$B$ Meson Decays}

\author{(CLEO Collaboration)}
\date{\today}
\maketitle
\tighten

\begin{abstract} 
      We search for CP-violating asymmetries (\acp ) in the 
      $B$ meson decays to
      $ K^\pm \pi^\mp$,
      $ K^\pm \pi^0$,
      $ \KS \pi^\pm$,
      $ K^\pm \etapr$, and
      $ \omega \pi^\pm$.  
      Using 9.66 million $\Upsilon (4S)$ decays collected with the
      CLEO detector, 
      the statistical precision on \acp\ is in the range of $\pm 0.12$ to
      $\pm 0.25$ depending on decay mode.
      While CP-violating asymmetries of up to $\pm 0.5$ are 
      possible within the Standard Model,
      the measured asymmetries
      are consistent with zero in all five decay modes studied.
\end{abstract}
\newpage

\renewcommand{\thefootnote}{\fnsymbol{footnote}}

\begin{center}
S.~Chen,$^{1}$ J.~Fast,$^{1}$ J.~W.~Hinson,$^{1}$ J.~Lee,$^{1}$
N.~Menon,$^{1}$ D.~H.~Miller,$^{1}$ E.~I.~Shibata,$^{1}$
I.~P.~J.~Shipsey,$^{1}$ V.~Pavlunin,$^{1}$
D.~Cronin-Hennessy,$^{2}$
% ...
\end{center}
 
\small
\begin{center}
$^{1}${Purdue University, West Lafayette, Indiana 47907}\\
$^{2}${University of Rochester, Rochester, New York 14627}\\
% ...
\end{center}

\setcounter{footnote}{0}
\newpage  
CP-violating phenomena arise in the Standard Model because of the
single complex parameter in the quark mixing matrix~\cite{km}. Such
phenomena are expected to occur widely in $B$ meson decays and will be 
searched for
by all current $B$-physics initiatives in the
world. However, there is currently no firm 
experimental evidence for CP violation outside the neutral kaon system,
where direct CP violation has been recently observed\cite{ktev}.  

Direct CP violation, i.e., a difference between the rates for
$\bar B\to \fbar$ and $B\to f $, will occur in any decay mode
for which there are two or more contributing amplitudes which differ
in both weak and strong phases. This rate difference gives rise to an 
asymmetry, \acp , defined as 
\begin{equation}
\acp\ \equiv 
 \frac{\branch(\bar B\to \fbar)-\branch(B\to f)}
{\branch(\bar B\to \fbar)+\branch( B\to f)}
.
\label{eqn:acpdef}
\end{equation}

For the simple case of two amplitudes $T,P$
with $T\ll P$, \acp\ is given by 
\begin{equation}
\acp\ \sim 2{|{T\over P}|} 
\sin\Delta\phi_{\mathrm{w}}\sin\Delta\phi_{\mathrm{s}}.
\label{eqn:acptheory}
\end{equation}
Here $\Delta\phi_{\mathrm{s}}$ and $\Delta\phi_{\mathrm{w}}$
refer to the difference in strong and weak phases between $T$ and $P$.

The decay $B\to K^\pm\pi^\mp$, for instance, involves a
$b\to u$ W-emission amplitude ($T$) with the
weak phase $\mathrm{Arg}(V_{ub}^*V_{us}^{})\equiv \gamma$ and a $b\to s$
penguin amplitude ($P$) with the weak phase 
$\mathrm{Arg}(V_{tb}^*V_{ts}^{}) = \pi$ or 
$\mathrm{Arg}(V_{cb}^*V_{cs}^{}) = 0$\cite{buras}.
Theoretical expectations of $|T/P|\sim 1/4 $ in $B\to K^\pm\pi^\mp$ 
~\cite{toverp}
thus allow for \acp\ as large as $\pm 0.5$.

The CP-violating phases may arise from either the Standard Model
CKM matrix or from new physics~\cite{nonsm}, while 
the CP-conserving strong phases
may arise from the absorptive part of a penguin diagram~\cite{BSS} 
or from final
state interaction effects~\cite{DGPS}. 
Precise predictions for \acp\ 
are not feasible at present as both the
absolute value and the strong interaction 
phases of the contributing amplitudes
are not calculable. 
However, numerical estimates can be made under well-defined model 
assumptions
and the dependence on both model parameters and
CKM parameters can be probed.
Recent calculations of CP
asymmetries under the assumption of factorization have been
published by Ali {\it et al.} \cite{cppred} and are listed in Table
\ref{tbl:results} for the modes examined in this paper.
A notable feature of the model used in Ref.~\cite{cppred} is that
soft final state interactions are neglected, leading to rather small
CP-invariant phases. However, it has been argued recently that 
CP-conserving
phases due to soft rescattering could be large~\cite{DGPS}, possibly 
leading to enhanced $|\acp|$~\cite{cplarge}. 

In 
this Letter, we present results of searches for
CP violation in decays of $B$ mesons to 
the three $K\pi$ modes, $ K^\pm \pi^\mp$, $
K^\pm \pi^0$, $ \KS \pi^\pm$, the mode $ K^\pm \etapr$, and the
vector-pseudoscalar mode $ \omega \pi^\pm$.  
These decay modes are selected
because they have well-measured branching ratios and significant
signal yields in our data sample~\cite{br}. In addition,
these decays are self-tagging; the flavor of the parent $b$ or
$\bar b$ quark is tagged simply by the sign of the high momentum
charged hadron. In the decay $B\to K^\pm\pi^\mp$ we assume 
that the charge of the
kaon tags the charge of the $b$ quark.

The data used in this analysis was collected with 
two configurations of the CLEO detector
at the Cornell Electron Storage Ring (CESR).  It
consists of an integrated luminosity of
$9.13~{\rm fb}^{-1}$ taken on the $\Upsilon$(4S)
resonance, corresponding to 9.66 million $B\bar{B}$ pairs, and $4.35~{\rm
fb}^{-1}$ taken below $B\bar{B}$ threshold, used for continuum
background studies.  
CLEO is a
general purpose solenoidal magnet detector,
described in detail elsewhere~\cite{detector}.  
Cylindrical drift
chambers in a 1.5T solenoidal magnetic field measure momenta and
specific ionization ($dE/dx$) of charged tracks. Photons are detected
using a 7800-crystal CsI(Tl) electromagnetic calorimeter.
For the second configuration, 
the innermost tracking chamber was replaced by a 3-layer, 
double-sided silicon
vertex detector, and the gas in the main drift chamber was changed
from an argon-ethane to a helium-propane mixture. These modifications
 led to improved
$dE/dx$ resolution in the main drift chamber, as well as improved
momentum resolution. Two thirds of the data used in the present analysis
was taken with the improved detector configuration.

Efficient track quality requirements are imposed on charged tracks.
Pions and kaons are identified by $dE/dx$.  
The separation between
kaons and pions for typical signal momenta $p \sim 2.6$~GeV$/c$\ is
$1.7$ and $2.0$ standard deviations ($\sigma$) for
the two detector configurations.
Candidate \KS\ are selected from pairs of tracks
forming well-measured displaced vertices with a $\pi^+\pi^-$
invariant mass within $2\sigma$ of the \KS\ mass. Pairs of
photons with an invariant mass within 2.5$\sigma$ of the nominal
$\pi^0$ mass are kinematically fitted with the mass
constrained to the nominal $\pi^0$ mass. 
For the high
momentum \KS\ and $\pi^0$ candidates reconstructed with these
requirements, the ratio of signal to combinatoric background is better
than 10.  
Electrons are rejected based on $dE/dx$ and the ratio of
the track momentum to the associated shower energy in the CsI
calorimeter; muons are rejected based on the penetration depth in
the instrumented steel flux return.  Resonances are reconstructed
through the decay channels $\eta'\to\eta\pi^+\pi^-$ with
$\eta\to\gamma\gamma$, $\eta'\to\rho\gamma$ with $\rho\to\pi^+\pi^-$,
and $\omega\to\pi^+\pi^-\pi^0$.

The \acp\ analyses presented here are closely related to the
corresponding branching fraction determinations published
elsewhere\cite{br}. We briefly summarize here the
main points of the analysis.

We calculate a beam-constrained $B$ mass $M = \sqrt{E_{\rm b}^2 -
p_B^2}$, where $p_B$ is the $B$ candidate momentum and $E_{\rm b}$ is
the beam energy. The resolution in $M$\ ranges from 2.5 to 3.0~${\rm
MeV} 
$, where the larger resolution corresponds to the
$B^\pm\to K^\pm\pi^0$ decay. We define $\Delta E = E_1 + E_2 - E_{\rm
b}$, where $E_1$ and $E_2$ are the energies of the daughters of the
$B$ meson candidate.  The resolution on $\Delta E$ is mode-dependent.
For final states without photons, the $\Delta E$ resolutions 
for the two configurations of the CLEO detector are
26 and 20 MeV. 
 We accept candidates with $M$\ within $5.2-5.3$~$\rm {GeV} 
$\
and $|\Delta E|<200$~MeV, and extract yields and asymmetries with
an unbinned maximum likelihood fit.  
The fiducial region in $M$ and $\Delta E$
includes the signal region and a substantial sideband for background
determination.  Sideband regions are also included around each of the
resonance masses ($\eta'$, $\eta$, and $\omega$) for use in the likelihood
fit.  For the $\eta'\to\rho\gamma$ case, the $\rho$ mass is not
included in the fit; we require $0.5\gev < m_{\pi\pi} < 0.9\gev$.

The main background arises from $e^+e^-\to q\bar q$\
(where $q=u,d,s,c$).  Such events typically exhibit a two-jet
structure and can produce high momentum back-to-back tracks in the
fiducial region.  To reduce contamination from these events, we
calculate the angle $\theta_{\rm s}$ between the sphericity 
axis\cite{shape}
of the candidate tracks and showers and the sphericity axis of the
rest of the event. The distribution of $\cossph$\ is strongly
peaked at $\pm 1$ for $q\bar q$\ events and is nearly flat for $B\bar
B$\ events. 
For $K\pi$ modes, 
we require $|\cossph|<0.8$\ which eliminates $83\%$\
of the background.  For \etapr\ and $\omega$ modes, the requirement is 
$|\cossph |< 0.9$.
%
detail in
%
Additional discrimination between signal and $q\bar q$ background is 
obtained 
from event shape information used in a Fisher discriminant (\fisher ) 
technique as described in detail in Ref.~\cite{bigrare}. 
%

Using a detailed GEANT-based Monte Carlo simulation~\cite{geant} we
determine overall detection efficiencies of 
0.48 ($K^\pm\pi^\mp$),
0.38 ($K^\pm\pi^0$),
0.15 ($K^0\pi^\pm$),
0.13 ($K^\pm\eta^\prime$), and
0.26 ($\omega\pi^\pm$).
These efficiencies
include secondary branching fractions for $K^0\to \KS\to \pi^+\pi^-$\
and $\pi^0\to \gamma\gamma$ as well as for the $\etapr$ and $\omega$ 
decay
modes where applicable.

To extract signal and background yields we perform unbinned
maximum-likelihood fits using $\Delta E$, $M$, ${\cal F}$,
$|\cos\theta_B|$ (if not used in ${\cal F}$),
$dE/dx$,
daughter resonance mass,
and helicity angle in the daughter decay.
The free parameters to be fitted are the 
asymmetry $(\fbar - f)/(\fbar + f)$ and the sum ($\fbar + f$)
in both signal and background. 
In most cases there is more than one possible signal hypothesis
and its corresponding background hypothesis,
e.g., 
we fit simultaneously for $K^\pm\pi^0$ and $\pi^\pm\pi^0$ to
ensure proper handling of the $K/\pi$ identification information. 
The probability density functions (PDFs)
describing the distribution of events in each variable are
parametrized by simple forms (Gaussian, polynomial, etc.) whose
parameters are determined in separate studies.  For signal PDF
shapes parameters are determined by fitting 
simulated signal events.  
Backgrounds in these analyses are dominated by
continuum \eeqq\ events, and we determine parameters of the background
PDFs by fitting data collected below the $\Upsilon(4S)$ resonance.
The uncertainties associated with such fits are
charge symmetric in all PDFs except the $dE/dx$ parametrization.
The $dE/dx$ information was calibrated 
such 
that 
any residual charge asymmetry is negligible compared to the
statistical errors for \acp .

%
%

The experimental determination of charge asymmetries in this analysis
depends entirely on the properties of high momentum tracks.  The
charged meson that tags the parent $b/\bar b$ flavor has in each case 
a momentum between 2.3 and 2.8~GeV/c.  In independent studies, using
very large samples of high momentum tracks, we searched for and
set stringent limits on the extent of possible charge-correlated bias
in the CLEO detector and analysis chain for tracks in the $2-3$~GeV$/c$ 
momentum range.  
Based on a sample of 8 million tracks, we find an \acp\ bias
of less than $\pm 0.002$
introduced by differences in reconstruction efficiencies for positive
and negative high momentum tracks.

For $K^\pm\pi^\mp$ combinations, where differential charge-correlated
efficiencies must also be considered in correlation with $K/\pi$
flavor, we use 37,000 $D^0\to K\pi(\pi^0)$ decays and set a
limit on the \acp\ bias of $\pm 0.005$. These $D^0$ meson decays,
together with an additional 24,000 $D^\pm_{(s)}$ meson decays, are also 
used
to set an upper limit of 0.4 MeV$/c$ on any charge-correlated or
charge-strangeness-correlated bias in the momentum measurement. The
resulting limit on \acp\ bias from this source is $\pm 0.002$.  We
conclude that there is no significant \acp\ bias introduced by track
reconstruction or selection.  

Our ability to distinguish the final states $K^+\pi^-$ and $K^-\pi^+$
depends entirely on particle identification using $dE/dx$. 
In addition, all other decay modes depend to varying degrees on $dE/dx$ to 
distinguish between $B\to X\pi^+$ and $X K^+$, $X$ being a 
$K^0_S, \pi^0, \eta^\prime$ or an $\omega$. 
The $dE/dx$ was
carefully calibrated in order to remove any possible charge dependencies.

\begin{figure}[hbp]
\centering
\leavevmode
\epsfxsize=3.25in
\epsffile{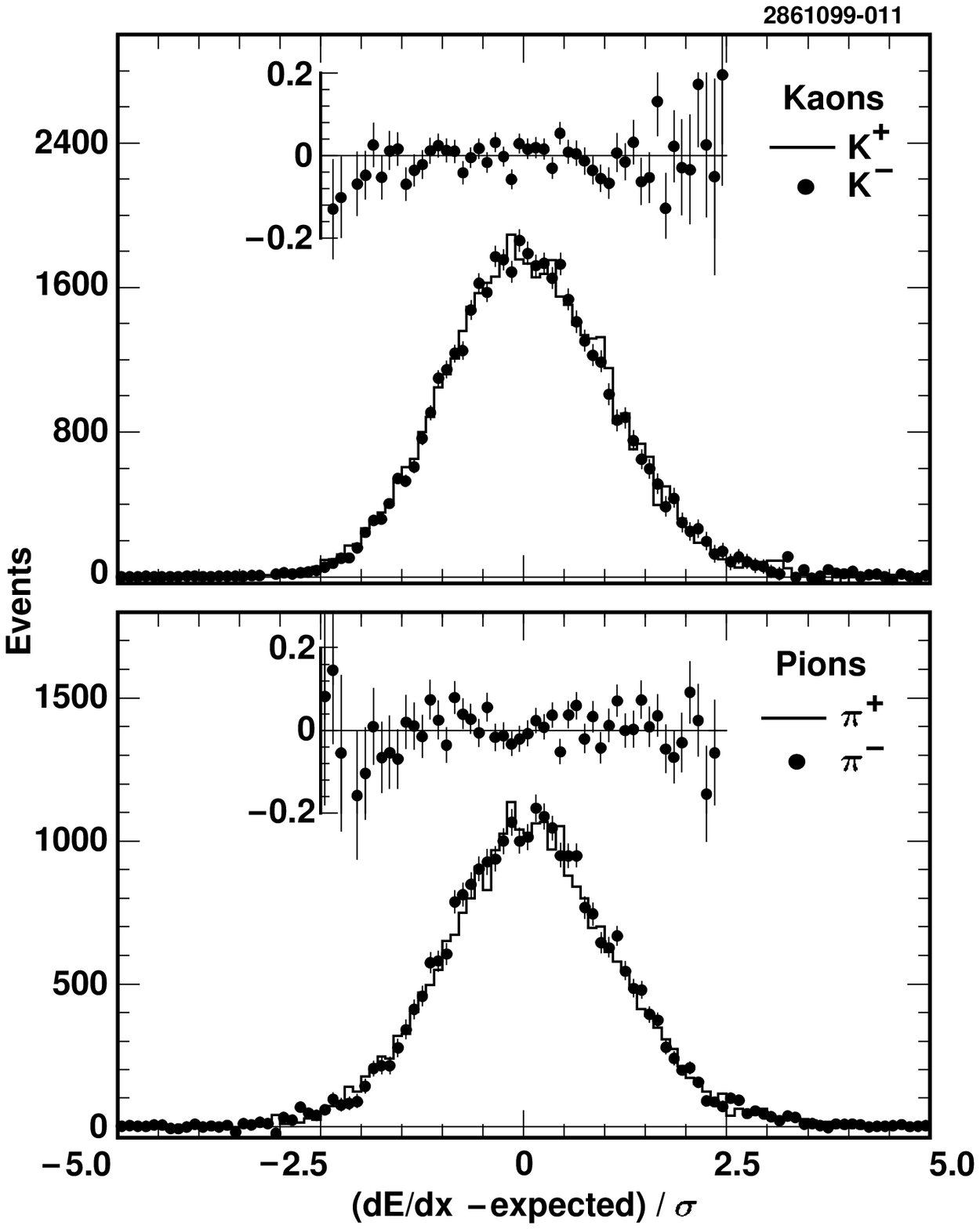}
\caption{Comparison of normalized \dedx\ distributions
for $K^+$ and $K^-$ (above); and
$\pi^+$ and $\pi^-$ (below). 
The inset shows the asymmetry 
$(K^- - K^+)/(K^- + K^+)$ and $(\pi^- - \pi^+)/(\pi^- + \pi^+)$ respectively.
Kinematically
identified kaons and pions 
of momenta $2-3$~GeV$/c$ from $D^0\to K\pi(\pi^0)$
decays in data are used for this comparison. }
\label{fig:dedx}
\end{figure}

We calibrate the $dE/dx$ response using radiative $\mu$ pair events
assuming that for a given velocity $dE/dx$ is the same 
for $\mu^\pm,\ \pi^\pm$, and $K^\pm$.
We then compare the dE/dx response for positive and negative 
tracks in the momentum range $2-3$~GeV$/c$ 
from all hadronic events in the CLEO data sample.
The large available statistics allows us to split the data into 
subsets and to 
verify the stability of the calibration
over time.
The fully calibrated $dE/dx$ is then verified using kinematically 
identified
kaons and pions of $2-3$~GeV$/c$ 
from $D^0\to K\pi(\pi^0)$ decays.
The $dE/dx$ distributions for 
$K^\pm$ and $\pi^\pm$ 
from this sample are shown
in Fig. \ref{fig:dedx}.
No significant
differences are seen between different charge species. 
The statistical uncertainty in this comparison translates into a possible
\acp\ bias of $\pm 0.01$ for $K^\pm\pi^\mp$, and less for all other
final states. 
We conservatively assign a total systematic error 
of $\pm 0.02$ in all five decay modes.

As additional check we measure the
asymmetry of the background events in each decay mode, and find that all
are consistent with the expected null result
for continuum background.
The results for the asymmetry in continuum background are
$-0.024\pm 0.038$ ($K^\pm\pi^\mp$), 
$-0.003\pm 0.032$ ($K^\pm\pi^0$),
$-0.017\pm 0.037$ ($K^0_S\pi^\pm$), 
$-0.006\pm 0.070$ ($\eta^\prime(\eta\pi\pi)K^\pm$), 
$-0.009\pm 0.015$ ($\eta^\prime(\rho\gamma)K^\pm$), 
and
$-0.001\pm 0.010$ ($\omega\pi^\pm$).
We further confirm that our analysis method does not introduce a bias 
in the measured \acp\ in the analysis of simulated events with known 
asymmetries.


We conclude that any possible systematic bias on \acp\ is negligible
compared to the statistical errors of our measurements.
Our 90$\%$ confidence level (CL)
ranges are calculated 
adding statistical
and systematic errors in quadrature.


\begin{table*}[ht]
\begin{center}
\caption{Summary of results. Signal yields are taken from
Ref. 10. Theory predictions are from 
Ref. 8, and include only Standard Model perturbative 
calculations.
The 90\% CL interval includes statistical and systematic errors 
($\pm 0.02$)
added in quadrature.}
\label{tbl:results}
\begin{tabular}{lcccc}
Mode 
& Signal & \acp\ & \acp\ & \acp\  \cr
&  Yield & & 90$\%$ CL & Theory \cr
\hline 
$K^\pm\pi^\mp$   
& $80^{+ 12}_{-11}$ 
& $-0.04\pm 0.16 $ 
& $[-0.30, 0.22]$ 
&  $ (+0.037, +0.106) $
\cr
$K^\pm \pi^0$    
& $42.1^{+10.9}_{-9.9}$ 
& $-0.29\pm 0.23 $ 
& $[-0.67, 0.09]$ 
& $ (+0.026, +0.092) $ 
\cr
$K^0_S\pi^\pm$     
&  $25.2^{+6.4}_{-5.6}$
& $+0.18\pm 0.24 $ 
&  $[-0.22, 0.56]$ 
& $ +0.015$ 
\cr
$K^\pm \eta^\prime$ 
& $100^{+13}_{-12}$
& $+0.03\pm 0.12 $  
& $[-0.17, 0.23]$ 
& $ (+0.020, +0.061) $ 
\cr
$\omega\pi^\pm$  
& $28.5^{+8.2}_{-7.3}$
& $-0.34\pm 0.25 $ 
&$[-0.75, 0.07]$ 
& $ (-0.120, +0.024)$ 
\cr
%
\end{tabular}
\end{center}
\end{table*}

%
%

\begin{figure}[hbp]
\centering
\leavevmode
\epsfxsize=3.25in
\epsffile{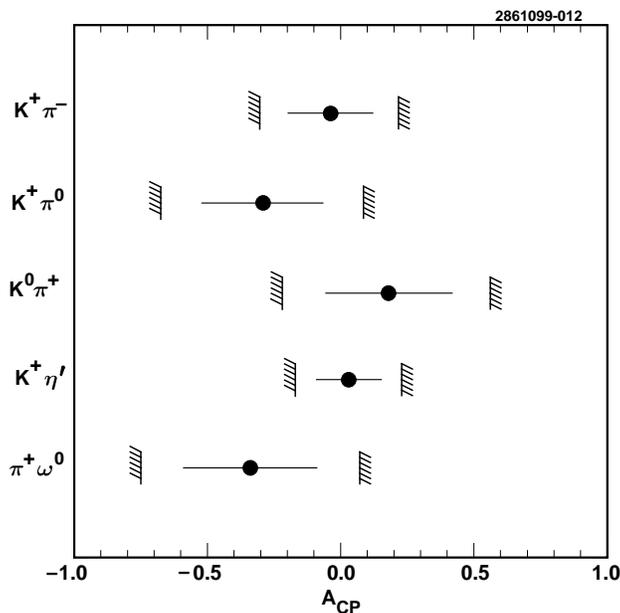}
\caption{\acp\ measurements. Error bars include systematics; hatched
regions delimit the 90\% CL intervals.}
\label{fig:prplot}
\end{figure}

\begin{figure}[hbp]
\centering
\leavevmode
\epsfxsize=3.25in
\epsffile{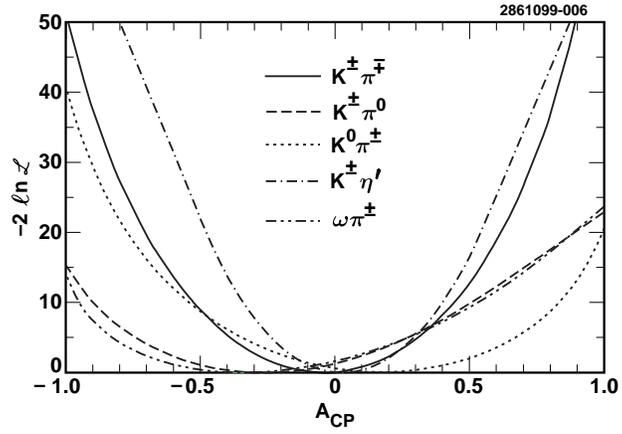}
\caption{Likelihood function ($-2\ln{\cal L}$) versus
\acp\ for each of the five modes.}
\label{fig:likelihoods}
\end{figure}

We summarize the results in
Table \ref{tbl:results} and Fig. \ref{fig:prplot}.
The dependence of the likelihood function on \acp\ for each of the five decay
modes is depicted in Fig.~\ref{fig:likelihoods}. 
This figure was obtained by re-optimizing the likelihood function
at each fixed value of \acp\ to account
for correlations between
the free parameters in the fit.

We see no evidence for CP
violation in the five modes analyzed here and set 90\% CL intervals
that reduce the possible range of \acp\ by as
much as a factor of four.  
It has been 
suggested~\cite{gronau-rosner} that \acp\ in 
$K^\pm\pi^\mp$ and $K^\pm\pi^0$ are expected to be almost identical 
within the Standard Model. 
Based on the average \acp\ in these two 
decay modes
we 
calculate
a $90\%$ CL range of $ -0.28 < $ \acp\ $< +0.05$.

While the sensitivity is not yet sufficient
to probe 
the rather small \acp\ values predicted by factorization
models, extremely large \acp\ values that might arise if large 
strong phase
differences were available from final state interactions are firmly 
ruled out. For
the cases of $K\pi$ and $\etapr K$, we can exclude 
$|\acp|$ greater than 0.30 and 0.23 at 90\% CL respectively.

We gratefully acknowledge the effort of the CESR staff in providing us
with excellent luminosity and running conditions. This work was
supported
 by the National Science
Foundation, the U.S. Department of Energy, the Research Corporation,
the Natural Sciences
and Engineering Research Council of Canada, the A.P.Sloan Foundation,
the Swiss National Science Foundation, and Alexander von Humboldt Stiftung.

\end{document}